\documentclass[aps,prb,bibnotes,twocolumn,showpacs,preprintnumbers,amsmath,amssymb,superscriptaddress,floatfix]{revtex4-2}
\usepackage[T1]{fontenc}
\usepackage[ansinew]{inputenc}
\usepackage{graphics}
\usepackage{dcolumn}
\usepackage{bm}
\usepackage{amsthm}
\usepackage{amsmath}
\usepackage{amssymb}
\usepackage{diagbox}
\usepackage{hyperref}
\usepackage{url}
\usepackage{xcolor}

\begin{document}

\title{Minimal model for vortex nucleation and reversal in spherical magnetic nanoparticles}

\def\afflux{Department of Physics and Materials Science, University of Luxembourg, 162A~Avenue de la Faiencerie, L-1511 Luxembourg, Grand Duchy of Luxembourg}

\author{Michael P.\ Adams}\email[Electronic address: ]{michael.adams@uni.lu}
\affiliation{\afflux}

\author{Andreas Michels}\email[Electronic address: ]{andreas.michels@uni.lu}
\affiliation{\afflux}


\begin{abstract}
Magnetic nanoparticles beyond the single-domain limit often develop vortex-like magnetization textures arising from the competition between exchange and magnetostatic energies. While such states are routinely studied using micromagnetic simulations, transparent analytical descriptions of vortex-mediated hysteresis and nucleation remain scarce. Here, we develop a semi-analytical minimal framework for vortex states in spherical magnetic nanoparticles. Guided by micromagnetic simulations, we introduce a parametrized vortex magnetization Ansatz based on hyperbolic functions that continuously interpolates between uniform and vortex states. In this way, we achieve a complexity reduction leading to a minimal Hamiltonian, which enables the efficient computation of magnetization curves and provides insight into vortex-mediated magnetization reversal. As an application, we derive analytical estimates for the critical vortex nucleation radius and field, recovering the functional form of Brown's classic result and extending it within a variational framework.
\end{abstract}

\date{\today}

\maketitle


\section{Introduction}

Magnetic nanoparticles constitute a paradigmatic system in which competing micromagnetic energy scales give rise to nontrivial magnetization textures, as demonstrated experimentally and theoretically across a wide range of length scales (see, e.g., Refs.~\cite{Usov2018_VortexHyperthermia,lappas2019,Vivas2020,BATLLE2022,Gerina2023,Niraula2023_VortexNanospheres,Adams2024,Sinaga2024,BATLLE2024,peddis2025} and references therein). While sufficiently small particles are well described by the single-domain Stoner-Wohlfarth model~\cite{Stoner1948}, larger particles reduce their magnetostatic energy by forming nonuniform configurations such as curling or vortex states. In spherical nanoparticles, vortex-like magnetization textures represent a particularly important class of flux-closure states and have therefore attracted sustained interest over several decades (see, e.g., Refs.~\cite{Niraula2023_VortexNanospheres,BATLLE2024} for recent experimental studies).

From a theoretical perspective, the description of vortex states in spherical magnetic nanoparticles has so far followed two largely separate routes. On the one hand, classic analytical works by Brown~\cite{brown1963micromagnetics} and Aharoni~\cite{aharoni2000introduction} provide elegant treatments of vortex nucleation based on fixed curling-type magnetization fields. These approaches yield closed-form estimates for the critical vortex-nucleation radius and offer valuable physical insight into the instability of the uniform state. However, they are inherently static in nature, as the assumed magnetization textures do not contain collective degrees of freedom that would allow for a continuous description of magnetization reversal or the emergence of hysteresis.

On the other hand, a large body of micromagnetic simulation studies using numerical frameworks such as \textsc{OOMMF}~\cite{oommf1999} or \textsc{MuMax3}~\cite{mumax2014} have explored vortex nucleation, noncentered vortex configurations, hysteresis loops, and dynamic effects for specific material systems. Although these simulations capture the full complexity and nonlinearity of the micromagnetic energy functional, their results are often strongly material- and discretization-dependent, and the underlying energetic mechanisms governing vortex-mediated reversal processes can be difficult to disentangle.

Despite the complementary nature of these two approaches, an intermediate-level theoretical description is missing that bridges static analytical theory and full micromagnetic simulations. Such a framework should ideally (i)~retain a small number of physically meaningful collective coordinates, (ii)~lead to an effective micromagnetic Hamiltonian amenable to analytical treatment, and (iii)~reproduce key features observed in experiments and simulations, including vortex-mediated hysteresis and analytical expressions for the nucleation field and radius.

Here, we introduce such a reduced description---a minimal model describing vortex nucleation and reversal in magnetic nanoparticles. Motivated by micromagnetic simulations of spherical nanoparticles~\cite{Adams2024vortex}, we construct a parametrized vortex magnetization Ansatz based on hyperbolic profile functions that captures the essential structure of the vortex state. The Ansatz contains a single variational parameter that controls the vortex-core width and allows for a global rotation of the magnetization texture, thereby enabling a {\it continuous} interpolation between the uniform and vortex states.

By inserting this parametrized magnetization field into the micromagnetic energy functional, we derive a reduced effective Hamiltonian that depends on only two collective coordinates. This reduced description provides a transparent account of vortex energetics, enables efficient computation of field-dependent magnetization curves, and allows for analytical estimates of the critical vortex-nucleation size and field. At the same time, it permits a critical comparison with full micromagnetic simulations and clarifies the physical origin of hysteresis features. In this way, the present framework establishes a direct and physically transparent link between analytical theory and contemporary micromagnetic simulations.

We refer to Supplemental Material~\cite{vortexsm2026} for additional micromagnetic calculations supporting the results of this work (see also Refs.~\cite{evans2014vampire,johnston2016magnetic,muller2019spirit,kronmuller2003micromagnetism} therein).

\begin{figure*}[tb!]
\centering
\resizebox{1.55\columnwidth}{!}{\includegraphics{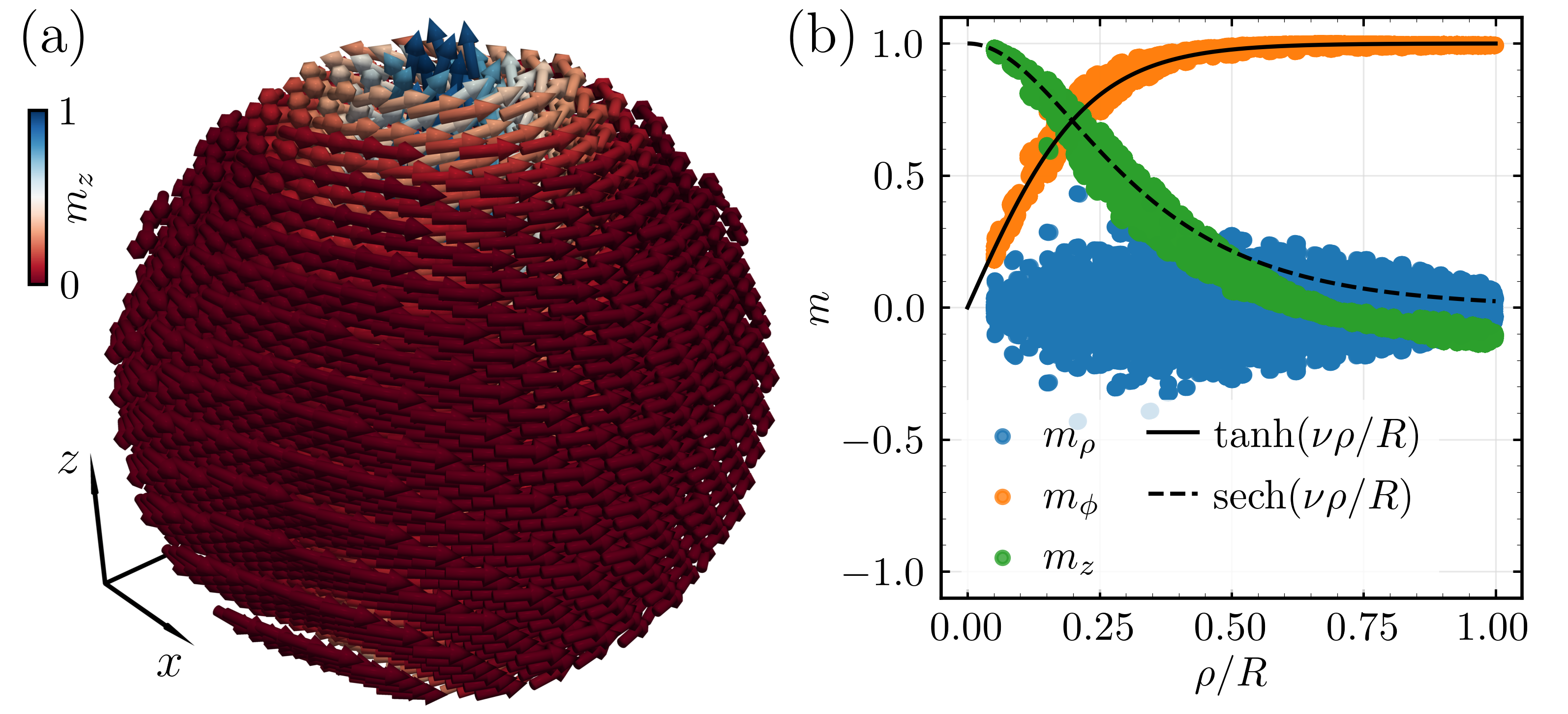}}
\caption{Micromagnetic simulation results (using \textsc{MuMax3}) for the remanent state of a spherical iron nanoparticle with a diameter of $D = 2R = 40 \, \mathrm{nm}$. The particle volume is discretized into cubical cells with a side length of $2 \, \mathrm{nm}$. (a)~Visualization of a stable vortex configuration; color encodes the $x$~component of the magnetization. (b)~Radial magnetization profiles obtained from several particles, shown in cylindrical coordinates $\{m_{\rho}, m_{\phi}, m_z\}$. The in-plane azimuthal component $m_{\phi}$ (solid line) is well described by a $\operatorname{tanh}$~profile, whereas the out-of-plane component $m_z$ (dashed line) follows a hyperbolic $\operatorname{sech}$~profile. Here, the vortex-profile parameter is $\nu = 4.443$.}
\label{fig1}
\end{figure*}

\section{Micromagnetic model}

We consider a spherical magnetic nanoparticle of radius $R$ described within the framework of continuum micromagnetics. The magnetization is represented by the unit vector field $\mathbf{m}(\mathbf{r})$, and the magnetic energy comprises exchange, uniaxial anisotropy, Zeeman, and magnetostatic contributions~\cite{Adams2024vortex}:
\begin{align}
\label{eq:efunc}
\mathcal{H}
&= A \int_V \sum_{i=x,y,z} (\nabla m_i)^2 \, d^3 r
- K_{\mathrm{u}} \int_V m_z^2 \, d^3 r \\
&\quad
- M_{\mathrm{s}} B_0 \int_V m_z \, d^3 r
- \frac{1}{2} M_{\mathrm{s}} \int_V \mathbf{m}\cdot\mathbf{B}_{\mathrm{d}} \, d^3 r \nonumber .
\end{align}
Here, $A$ denotes the exchange-stiffness constant, $K_{\mathrm{u}} > 0$ is the first-order uniaxial anisotropy constant, and $M_{\mathrm{s}}$ is the saturation magnetization. The magnetization vector field is given by $\mathbf{M}(\mathbf{r}) = M_{\mathrm{s}} \mathbf{m}(\mathbf{r})$, the (constant) externally applied magnetic field is written as $\mathbf{B}_0 = \mu_0 \mathbf{H}_0 = B_0 \mathbf{e}_z$, where $\mathbf{H}_0$ denotes the magnetic field strength, $\nabla = \{ \partial/\partial x, \partial/\partial y, \partial/\partial z \}$, and all integrals extend over the volume $V$ of the particle. The demagnetizing field $\mathbf{B}_{\mathrm{d}} = \mu_0 \mathbf{H}_{\mathrm{d}}$ is obtained from the standard magnetostatic convolution with the demagnetization kernel, where $\mu_0 = 4\pi \times 10^{-7} \, \mathrm{Tm/A}$.

Throughout this work, both the uniaxial anisotropy axis and the external magnetic field are chosen to be parallel to the $z$~axis of a Cartesian laboratory frame. This configuration isolates the essential energetics of vortex-like magnetization states while maintaining analytical transparency. In all numerical simulations and analytical calculations, the following material parameters for iron were used (for simplicity, we assumed a uniaxial rather than a cubic anisotropy): $K_{\mathrm{u}} = 4.8 \times 10^4 \, \mathrm{J/m}^3$, $M_{\mathrm{s}} = 1700 \, \mathrm{kA/m}$, and $A = 10 \, \mathrm{pJ/m}$. The applied field $B_0$ is varied within the range $\pm 1 \, \mathrm{T}$ for hysteresis simulations, while the particle radius $R$ is varied between $8$ and $20 \, \mathrm{nm}$ to investigate the vortex-nucleation size.

\section{Hyperbolic vortex Ansatz}

Numerical micromagnetic simulations of spherical iron nanoparticles reveal that the remanent magnetization state ($B_0 = 0 \,\mathrm{T}$) is characterized by a vortex-like texture with a smooth out-of-plane core and an azimuthal in-plane magnetization component~\cite{Adams2024vortex}. As shown in Fig.~\ref{fig1}, when expressed in cylindrical coordinates, the radial magnetization component $m_{\rho}$ vanishes on average, while the tangential component $m_{\phi}$ increases towards unity due to magnetic flux closure and the out-of-plane component $m_z$ exhibits a smooth decay from the particle center towards the surface. Remarkably, these radial profiles collapse onto simple hyperbolic functional forms [solid and dashed lines in Fig.~\ref{fig1}(b)].

Motivated by the appearance of these hyperbolic profiles, we consider an effective {\it local} vortex Ansatz that captures the essential structure of the spin texture:
\begin{align}
\label{ansatz1}
&\mathbf{m}'(\rho,\phi,z;\nu) = \\
&\left\{
-\tanh(\nu \rho/R) \sin\phi ,\,
\tanh(\nu \rho/R) \cos\phi ,\,
\operatorname{sech}(\nu \rho/R)
\right\} \nonumber ,
\end{align}
which satisfies $\|\mathbf{m}'\| = 1$ identically. The parameter $\nu$ controls the width of the vortex core and continuously interpolates between the uniform state ($\nu = 0$) and a fully developed vortex ($\nu \gg 1$). To introduce degrees of freedom that allow for magnetization reversal, we express the {\it global} magnetization $\mathbf{m}$ as the rotation of $\mathbf{m}'$ by two angles $\omega$ and $\tau$:
\begin{equation}
\label{ansatz2}
\mathbf{m}(\rho,\phi,z; \omega,\tau; \nu)
= \mathbf{R}(\omega,\tau)\cdot \mathbf{m}'(\rho,\phi,z; \nu) ,
\end{equation}
where the rotation matrix $\mathbf{R}(\omega, \tau) = \mathbf{R}_z(\omega) \cdot \mathbf{R}_y(\tau)$ is defined as the product of a rotation about the $z$~axis and a rotation about the $y$~axis~\cite{vortexsm2026}. The angle $\tau$ describes the inclination of the vortex axis with respect to the field direction and $\omega$ specifies a rotation in the $x$-$y$-plane.

The hyperbolic vortex Ansatz provides a compact and physically motivated parameterization of the remanent (low field) magnetization texture in terms of a small set of collective variables. In the following, this Ansatz is used to derive a reduced effective Hamiltonian by explicit insertion into the micromagnetic energy functional~\eqref{eq:efunc}.

\section{Reduced effective Hamiltonian}

Inserting the hyperbolic vortex Ansatz defined by Eqs.~(\ref{ansatz1}) and (\ref{ansatz2}) into the micromagnetic energy functional~\eqref{eq:efunc} yields a reduced effective description in terms of a small set of collective variables, with the individual energy contributions expressed through dimensionless profile functions $g_i(\nu)$. The resulting reduced Hamiltonian $\mathcal{H}'$ reads (see \cite{vortexsm2026} for details):
\begin{align}
\label{eq:ReducedHamiltonian1}
&\mathcal{H}'(\nu,\tau, B_0) = \\ 
&g_{\mathrm{ex}}(\nu) - \frac{K_{\mathrm{u}} R^2}{A} \left[ g_{\mathrm{u}}^{x}(\nu)\sin^2\tau + g_{\mathrm{u}}^{z}(\nu)\cos^2\tau \right] \nonumber \\
&- \frac{M_{\mathrm{s}} R^2 B_0}{A} g_{\mathrm{z}}^{z}(\nu)\cos\tau
- \frac{\mu_0 M_{\mathrm{s}}^2 R^2}{A} g_{\mathrm{d}}(\nu) \nonumber .
\end{align}
The exchange ($g_{\mathrm{ex}}$), anisotropy ($g_{\mathrm{u}}^{x}$ and $g_{\mathrm{u}}^{z}$), and Zeeman ($g_{\mathrm{z}}$) contributions reduce to two-dimensional integrals over the unit sphere, while the evaluation of the magnetodipolar term ($g_{\mathrm{d}}$), requiring the solution of a six-fold integral, is accomplished via a numerical grid scheme. Note that the anisotropy contribution to $\mathcal{H}'$ consists of a term $g_{\mathrm{u}}^{x} \sin^2\tau$ that prefers the in-plane orientation and a term $g_{\mathrm{u}}^{z} \cos^2\tau$ that favors the collective out-of-plane orientation of the magnetization.

We note that the reduced Hamiltonian $\mathcal{H}'$ is independent of the azimuthal angle $\omega$. This invariance directly follows from the rotational symmetry of the underlying micromagnetic Hamiltonian $\mathcal{H}$ [Eq.~\eqref{eq:efunc}] for a uniaxial anisotropy aligned with the applied magnetic field. If the anisotropy axis is tilted with respect to the external field, this symmetry is broken and an explicit $\omega$~dependence of $\mathcal{H}'$ would generally remain.

\section{Evaluation of the reduced Hamiltonian $\mathcal{H}'$}

In the following, we discuss the evaluation of the Hamiltonian $\mathcal{H}'(\nu, \tau, B_0)$ by direct comparison with micromagnetic simulations performed with \textsc{MuMax3}, based on the original Hamiltonian $\mathcal{H}(\mathbf{m})$ [Eq.~(\ref{eq:efunc})]. For the minimization of $\mathcal{H}'(\nu, \tau, B_0)$, the profile functions $g_i(\nu)$ were first numerically evaluated by direct integration for the parameter range $\nu \in [0,10]$~\cite{vortexsm2026}. To speed up the energy minimization, the resulting data were stored as lookup tables and reused in a field-following procedure for a magnetic field cycle starting from $B_0 = 1 \, \mathrm{T}$ and for a total number of $1000$ field steps. 

\begin{figure*}[tb!]
\centering
\resizebox{1.55\columnwidth}{!}{\includegraphics{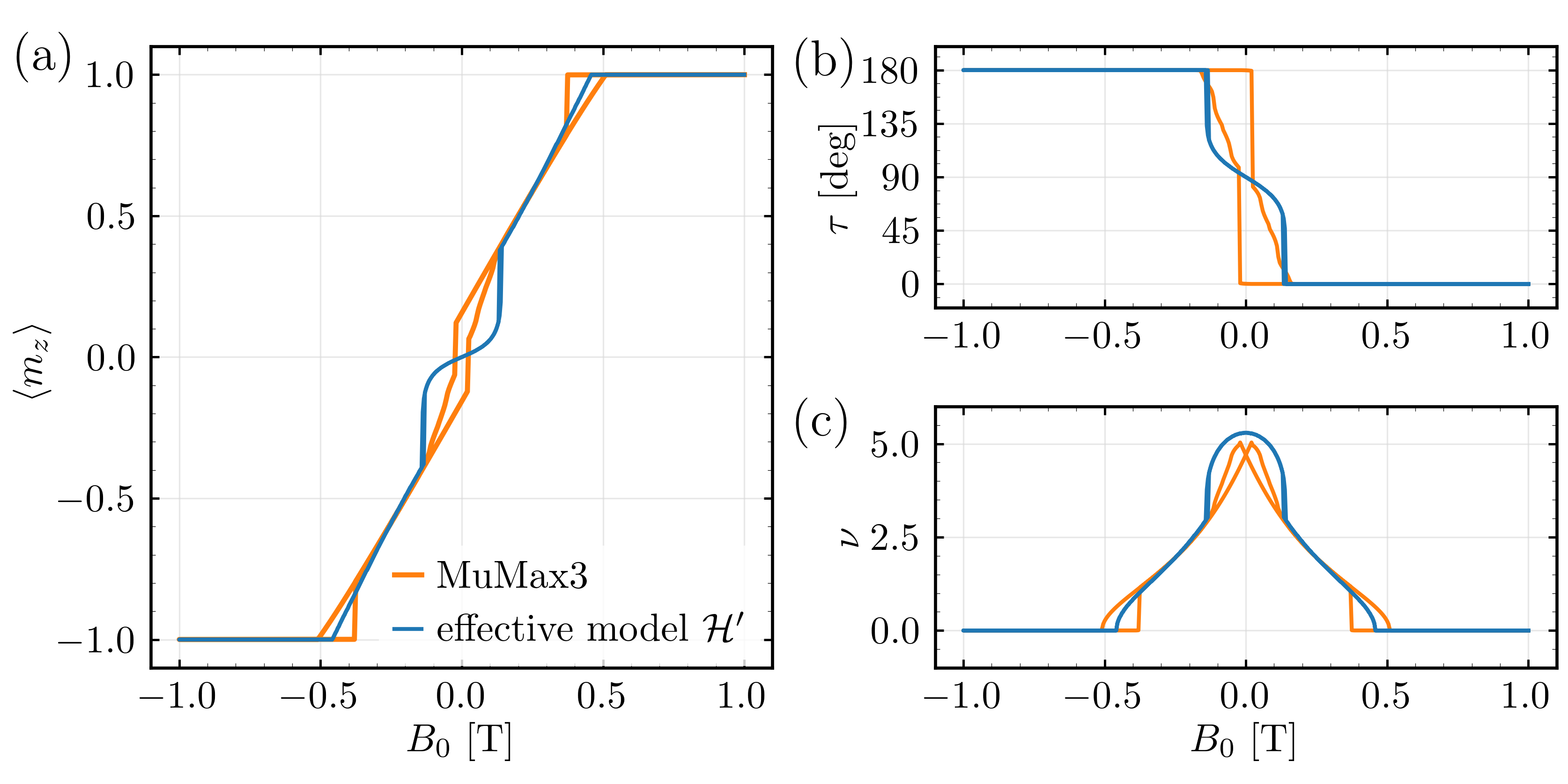}}
\caption{Comparison between the semi-analytical modeling approach (solid blue curves) and the micromagnetic simulation results obtained with \textsc{MuMax3} (solid orange curves) for a spherical magnetic nanoparticle in the vortex state. The magnetization response is computed from the effective Hamiltonian $\mathcal{H}'$ [Eq.~\eqref{eq:ReducedHamiltonian1}] by minimizing the total energy with respect to the vortex profile parameter $\nu$ and the inclination angle $\tau$ at each applied field value $B_0$. The right-hand panels show the corresponding field-dependent evolution of $\tau$~(b) and $\nu$~(c). The \textsc{MuMax3} simulation was carried out for a spherical particle of diameter $D = 40 \, \mathrm{nm}$, using a cubic discretization with a cell side length of $2 \, \mathrm{nm}$.}
\label{fig2}
\end{figure*}

Figure~\ref{fig2}(a) compares the magnetization curves obtained from the reduced Hamiltonian $\mathcal{H}'$ with the corresponding micromagnetic simulation result, and Figs.~\ref{fig2}(b) and (c) show the dependence of the angle $\tau$ and of the vortex-core parameter $\nu$ on the applied magnetic field $B_0$. Here, the average magnetization $\langle m_z \rangle$ is defined by the following integral, which is expressible as a function of $\nu$ and $\tau$:
\begin{align}
    \langle m_z \rangle &= \frac{1}{V} \int_{V} m_z(\mathbf{r}) \, d^3r = g_{\mathrm{z}}^{z}(\nu) \cos\tau .
\end{align}
Starting on the upper branch in Fig.~\ref{fig2}(a), at $B_0 = 1 \, \mathrm{T}$, both approaches are in the saturated regime ($\langle m_z \rangle = 1$) down to approximately $B_0 = 0.5 \, \mathrm{T}$. Below $B_0 = 0.5 \, \mathrm{T}$ a deviation between the two approaches is observed: the effective Hamiltonian $\mathcal{H}'$ produces a continuous transition with an approximately linear decay of $\langle  m_z \rangle$, whereas the \textsc{MuMax3} result exhibits a jump at $B_0 \cong 0.36 \, \mathrm{T}$. Following this transition, the two magnetization curves approach each other again and remain close until $B_0 \simeq 0.12 \, \mathrm{T}$. In the interval $-0.12 \, \mathrm{T} \le B_0 \le 0.12 \, \mathrm{T}$, the behavior differs qualitatively. While the \textsc{MuMax3} simulation exhibits hysteretic behavior with finite remanent magnetization and a coercive field, the reduced Hamiltonian $\mathcal H'$ follows a smooth and reversible path with vanishing net magnetization at $B_0 = 0$. Similarly, the field dependencies of $\tau$ and $\nu$ [Figs.~\ref{fig2}(b) and (c)] also reveal a smooth transition for $\mathcal{H}'$. For smaller applied fields below $-0.12 \, \mathrm{T}$, the magnetization behavior is symmetric to the upper branch.

\begin{figure*}[tb!]
\centering
\resizebox{1.55\columnwidth}{!}{\includegraphics{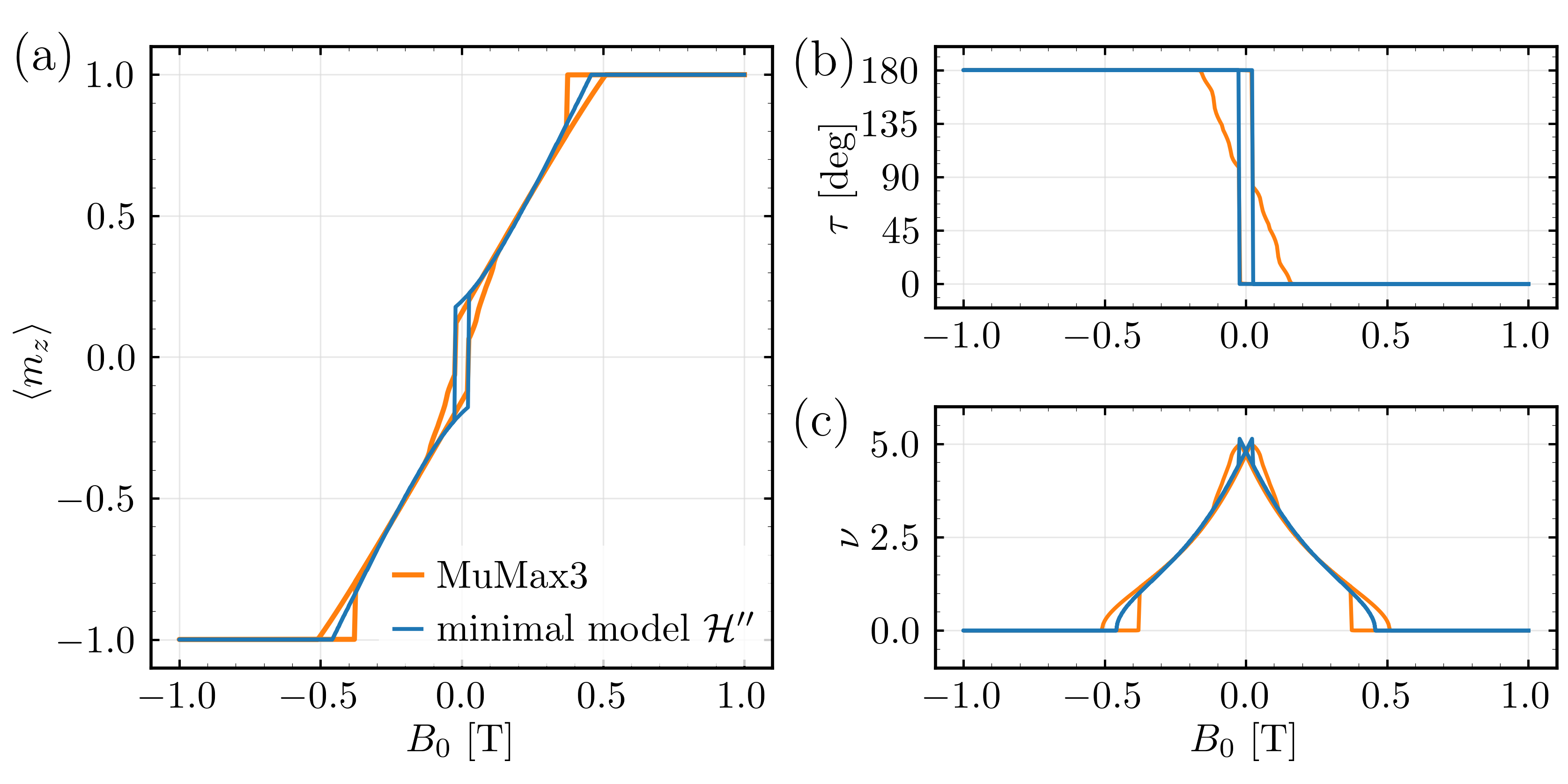}}
\caption{Same as Fig.~\ref{fig2}, but with the minimal-model Hamiltonian $\mathcal{H}''$ defined by Eq.~\eqref{eq:VortexReducedSimplifiedHamiltonian}.}
\label{fig3}
\end{figure*}

Inspection of the reduced Hamiltonian $\mathcal{H}'$ [Eq.~\eqref{eq:ReducedHamiltonian1}] shows that this nonhysteretic reversal path is driven by the anisotropy contribution proportional to $\sin^2\tau$, which is the only term in $\mathcal{H}'$ that energetically favors an in-plane magnetization orientation. As a consequence, the reduced Hamiltonian $\mathcal{H}'$ captures only a single continuous reversal path and does not reproduce the metastable behavior required for hysteretic magnetization reversal.

\section{Introduction of the minimal Hamiltonian $\mathcal{H}''$}

The shortcoming of $\mathcal H'$ revealed by the above comparison motivates the introduction of the Hamiltonian $\mathcal H''$:
\begin{align}
    \mathcal{H}''(\nu, \tau, B_0) &= g_{\mathrm{ex}}(\nu) - \frac{K_{\mathrm{u}} R^2}{A} \, g_{\mathrm{u}}^{z}(\nu) \cos^2\tau 
    \label{eq:VortexReducedSimplifiedHamiltonian}
    \\
    &- \frac{M_{\mathrm{s}} R^2 B_0}{A} \, g_{\mathrm{z}}^{z}(\nu) \cos\tau - \frac{\mu_0 M_{\mathrm{s}}^2 R^2}{A} \, g_{\mathrm{d}}(\nu) \nonumber ,
\end{align}
where the anisotropy contribution proportional to $\sin^2\tau$ is omitted, as it enforces a nonhysteretic reversal path in the reduced energy landscape. Figure~\ref{fig3}(a) compares the magnetization curves obtained from the Hamiltonian $\mathcal{H}''$ with the corresponding \textsc{MuMax3} simulation result. We find that $\mathcal{H}''$ and $\mathcal{H}'$ (cf.\ Fig.~\ref{fig2}) yield equivalent collective magnetization behavior for applied fields $|B_0| \ge 0.12 \, \mathrm{T}$. In contrast to $\mathcal{H}'$, $\mathcal{H}''$ exhibits, however, a hysteresis loop for $|B_0| \le 0.12 \, \mathrm{T}$, in quantitative agreement with the micromagnetic simulation. Furthermore, in contrast to Figs.~\ref{fig2}(b) and (c), the field dependence of $\tau$ and $\nu$ in Fig.~\ref{fig3}(b) and (c) becomes hysteretic. This indicates that the magnetization reversal proceeds via competing metastable vortex configurations, rather than along a single smooth rotational path in the reduced phase space. The effective Hamiltonian $\mathcal{H}''$ retains the relevant vortex degrees of freedom while allowing for metastable vortex configurations and hysteretic magnetization reversal.

Finally, we note that the remaining discrepancy between the effective model $\mathcal{H}''$ and the \textsc{MuMax3} simulation in Fig.~\ref{fig3}(a) is highly sensitive to the discretization cell size employed in the micromagnetic calculation ($2 \times 2 \times 2 \, \mathrm{nm}^3$ in Figs.~\ref{fig2} and \ref{fig3})~\cite{vortexsm2026}. This sensitivity indicates that the observed difference is influenced by systematic discretization effects inherent to the finite-difference micromagnetic model (compare also to Ref.~\cite{Holt2025_discretization}). As a consequence, it cannot be unambiguously decided from this comparison alone whether the effective continuum model or the discretized simulation more closely represents the continuum-limit behavior.

\section{Stability properties and classical limits of $\mathcal H''$}

Here, we demonstrate that the reduced vortex Hamiltonian $\mathcal H''(\nu, \tau, B_0)$ provides a unified variational description of both vortex nucleation and coherent single-domain reversal. All results discussed below follow from the stability properties of $\mathcal H''$ [Eq.~\eqref{eq:VortexReducedSimplifiedHamiltonian}] and require no additional assumptions.

\begin{figure}[tb!]
\centering
\resizebox{0.85\columnwidth}{!}{\includegraphics{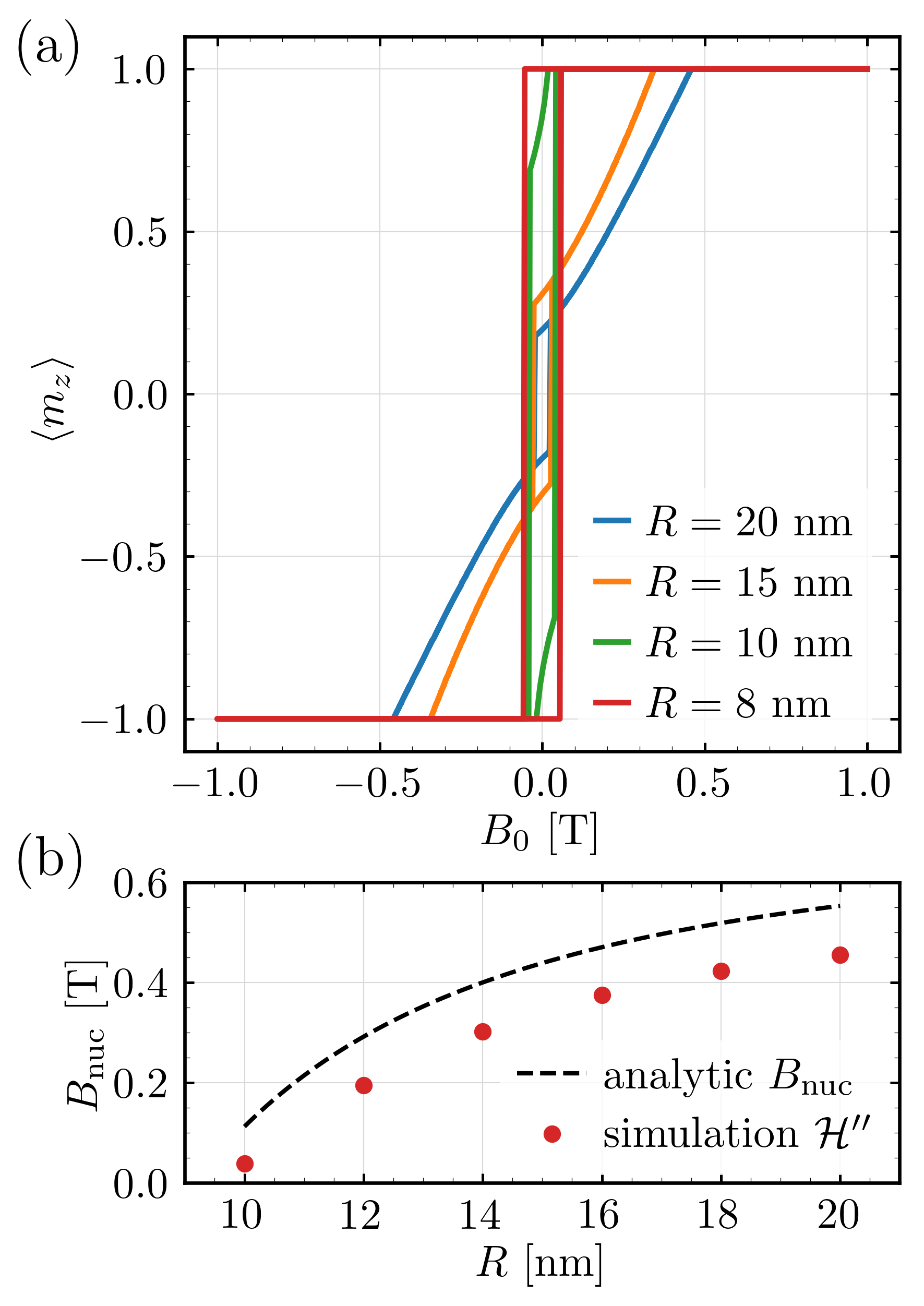}}
\caption{Magnetization reversal and vortex nucleation obtained from the minimal vortex Hamiltonian $\mathcal{H}''$ [Eq.~\eqref{eq:VortexReducedSimplifiedHamiltonian}]. (a)~Field-dependent average magnetization $\langle m_z \rangle$ for different particle radii $R$ (see inset). For $R < R_{\mathrm{nuc}} \cong 9.2 \,\mathrm{nm}$ [Eq.~\eqref{eq:VortexNucleationRadius}], the system exhibits single-domain Stoner-Wohlfarth hysteresis (rectangular loop for $R = 8 \,\mathrm{nm}$) with abrupt switching at $B_{\mathrm{SW}} = 2 K_{\mathrm{u}}/M_{\mathrm{s}} \cong 0.056 \, \mathrm{T}$. For $R > R_{\mathrm{nuc}}$, vortex formation enables a smooth, nonuniform magnetization reversal. (c)~Vortex-nucleation field $B_{\mathrm{nuc}}$ versus particle radius $R$. The analytical expression~\eqref{eq:VortexNucleationField} (dashed line) corresponds to the linear stability (spinodal) limit, while the hysteresis data reflect the loss of metastability along the reversal path.}
\label{fig4}
\end{figure}

The onset of vortex formation is governed by the loss of stability of the homogeneous state with respect to infinitesimal vortex distortions ($\nu \to 0$). Evaluating the corresponding stability condition,
$\partial^2 \mathcal H''/\partial \nu^2 = 0$~\cite{vortexsm2026}, yields the critical vortex-nucleation radius ($B_0 = 0 \, \mathrm{T}$):
\begin{equation}
R_{\mathrm{nuc}} \cong \frac{42\sqrt{5}}{25} \, \ell_{\mathrm{ex}}
\left[
1 - \frac{3528}{625}\frac{K_{\mathrm{u}}}{\mu_0 M_{\mathrm{s}}^2}
\right]^{-1/2},
\label{eq:VortexNucleationRadius}
\end{equation}
where $\ell_{\mathrm{ex}} = \sqrt{2A / (\mu_0 M_{\mathrm{s}}^2)}$ denotes the magnetostatic exchange length. Equation~(\ref{eq:VortexNucleationRadius}) reproduces the functional form of Brown's classic vortex-nucleation criterion~\cite{brown1968fundamental,aharoni2000introduction}, with small quantitative differences arising from the variational Ansatz employed here. The same stability condition can equivalently be expressed in terms of the applied magnetic field, yielding the corresponding vortex-nucleation field ($R > R_{\mathrm{nuc}}$):
\begin{equation}
B_{\mathrm{nuc}} = \mu_0 H_{\mathrm{nuc}} = 
\frac{625}{1764} \mu_0 M_{\mathrm{s}}
- \frac{2K_{\mathrm{u}}}{M_{\mathrm{s}}}
- \frac{10A}{M_{\mathrm{s}} R^2}.
\label{eq:VortexNucleationField}
\end{equation}
Equations~\eqref{eq:VortexNucleationRadius} and~\eqref{eq:VortexNucleationField} therefore represent two equivalent parametrizations of the same spinodal stability limit of the homogeneous state.

For particle radii $R$ below $R_{\mathrm{nuc}}$, the vortex degree of freedom is suppressed ($\nu = 0$), and the Hamiltonian $\mathcal H''$ reduces identically to the uniaxial macrospin model. In this limit, the Stoner-Wohlfarth switching field, $B_{\mathrm{SW}} = 2K_{\mathrm{u}} / M_{\mathrm{s}}$, is recovered without further assumptions~\cite{vortexsm2026}. Figure~\ref{fig4}(a) illustrates the resulting crossover between coherent rotation and vortex-mediated magnetization reversal and highlights the continuous nature of the transition within a single variational framework. The analytical prediction for $B_{\mathrm{nuc}}$ [Eq.~(\ref{eq:VortexNucleationField})] exhibits a constant offset relative to the nucleation fields that are numerically obtained using $\mathcal{H}''$ [Fig.~\ref{fig4}(b)], which is expected as Eq.~\eqref{eq:VortexNucleationField} follows from a local instability analysis.

\section{Conclusion}

We have developed a reduced semi-analytical description of vortex states in spherical magnetic nanoparticles based on a hyperbolic magnetization Ansatz. The resulting effective Hamiltonian captures the essential energetics of vortex formation and reversal while remaining analytically transparent and computationally efficient. The framework bridges the gap between large-scale micromagnetic simulations and analytical theory, provides physical insight into hysteresis mechanisms and vortex nucleation, and highlights the influence of numerical discretization on simulated magnetization curves. As such, it offers a useful reference model for both theoretical studies and the interpretation of experimental nanoparticle data in nanomagnetism.

The continuously-tunable hyperbolic vortex profile employed in this work does not emerge in an obvious way from the micromagnetic Hamiltonian alone. Instead, it is revealed empirically by treating micromagnetic simulations as numerical experiments and subsequently identifying a functional form that robustly captures the observed magnetization profiles. The resulting reduced description is therefore not a direct analytical solution of the Hamiltonian but an effective minimal model constructed from simulation-informed physical insight. It is of interest to extend this study to other particle shapes and to arbitrary orientation of the anisotropy axis with respect to the applied field. Likewise, in addition to the average magnetization, one may use the present approach to analytically compute other experimental observables, such as the magnetic small-angle neutron scattering cross section.

{\it Data availability}---Numerical data and codes (MuMax3 and python scripts) corresponding to this study are available on Zenodo~\cite{Adams2026_Vortex}.


%

\end{document}


\title{Supplemental Material to ``Minimal model for vortex nucleation and reversal in spherical magnetic nanoparticles''}

\def\afflux{Department of Physics and Materials Science, University of Luxembourg, 162A~Avenue de la Faiencerie, L-1511 Luxembourg, Grand Duchy of Luxembourg}

\author{Michael P.\ Adams}\email[Electronic address: ]{michael.adams@uni.lu}
\affiliation{\afflux}

\author{Andreas Michels}\email[Electronic address: ]{andreas.michels@uni.lu}
\affiliation{\afflux}


\begin{abstract}
We provide additional analytical micromagnetic results to the above article. Moreover, a video file features the magnetization reversal process as obtained from the numerical micromagnetic simulations.
\end{abstract}

\date{\today}

\maketitle


\section{Derivation of the reduced Hamiltonian}
\label{appa}

We consider the following micromagnetic Hamiltonian for a spherical nanoparticle of radius $R$:
\begin{align}
\label{eq:efuncAppx}
\mathcal{H}
&= A \int_V \sum_{i=x,y,z} (\nabla m_i)^2 \, d^3 r
- K_{\mathrm{u}} \int_V m_z^2 \, d^3 r \\
&\quad
- M_{\mathrm{s}} B_0 \int_V m_z \, d^3 r
- \frac{1}{2} M_{\mathrm{s}} \int_V \mathbf{m}\cdot\mathbf{B}_{\mathrm{d}} \, d^3 r \nonumber ,
\end{align}
where $A$ denotes the exchange-stiffness constant, $K_{\mathrm{u}} > 0$ is the first-order uniaxial anisotropy constant, and $M_{\mathrm{s}}$ is the saturation magnetization. The magnetization vector field is given by $\mathbf{M}(\mathbf{r}) = M_{\mathrm{s}} \, \mathbf{m}(\mathbf{r})$, the (constant) externally applied magnetic field is written as $\mathbf{B}_0 = \mu_0 \mathbf{H}_0 = B_0 \mathbf{e}_z$, where $\mathbf{H}_0$ denotes the magnetic field strength, $\nabla = \{ \partial/\partial x, \partial/\partial y, \partial/\partial z \}$, and all integrals extend over the volume $V$ of the particle. The demagnetizing field $\mathbf{B}_{\mathrm{d}} = \mu_0 \mathbf{H}_{\mathrm{d}}$ is obtained from the standard magnetostatic convolution with the demagnetization kernel, where $\mu_0 = 4\pi \times 10^{-7} \, \mathrm{Tm/A}$.

To reduce model complexity and the degrees of freedom, we express the local unit magnetization vector field $(\|\mathbf{m}'\| = 1)$ as follows
\begin{equation}
\label{ansatz1Appx}
\mathbf{m}'(\rho,\phi,z;\nu)
=
\left\{
-\tanh(\nu \rho/R)\sin\phi,\;
\tanh(\nu \rho/R)\cos\phi,\;
\operatorname{sech}(\nu \rho/R)
\right\} .
\end{equation}
Equation~(\ref{ansatz1Appx}) is motivated from micromagnetic simulations using \textsc{MuMax3}~\cite{Adams2024vortex}, where $\nu$ controls the width of the vortex core ($\nu = 0$ for the uniform state and $\nu \gg 1$ for a fully developed vortex). To allow for magnetization reversal, we further assume the global magnetization vector field $\mathbf{m}$ as a rotation of $\mathbf{m}'$ by two angles $\omega,\;\tau$:
\begin{align}
\mathbf{m}(\rho, \phi, z; \omega, \tau; \nu) = \mathbf{R}(\omega, \tau) \cdot \mathbf{m}'(\rho, \phi, z; \nu),
\label{eq:VortexAnsatzAppx}
\end{align}
with the rotation matrix
\begin{align}
\mathbf{R}(\omega, \tau) = \mathbf{R}_z(\omega)\cdot\mathbf{R}_y(\tau) =
\begin{bmatrix} 
\cos\omega \cos\tau & -\sin\omega & \cos\omega \sin\tau  \\
\sin\omega \cos\tau & \cos\omega & \sin\omega \sin\tau \\
- \sin\tau & 0  & \cos\tau
\end{bmatrix} .
\label{eq:VortexParticlesRotationMatrixapp}
\end{align}
Inserting the Ansatz [Eq.~\eqref{eq:VortexAnsatzAppx}] into the energy functional [Eq.~\eqref{eq:efuncAppx}] yields compact expressions for the individual energy contributions in terms of dimensionless profile integrals $g_i(\nu)$:
\begin{align}
\mathcal{H}_{\mathrm{ex}}(\nu) &= A \int_V \sum_{i=x,y,z} (\nabla m_i)^2 \, d^3 r= \frac{4\pi A R}{3} \, g_{\mathrm{ex}}(\nu), \\
\mathcal{H}_{\mathrm{u}}(\nu,\tau) &=- K_{\mathrm{u}} \int_V m_z^2 \, d^3 r =- \frac{4\pi R^3 K_{\mathrm{u}}}{3} \left(g_{\mathrm{u}}^{x}(\nu) \sin^2\tau + g_{\mathrm{u}}^{z}(\nu) \cos^2\tau\right), \\
\mathcal{H}_{\mathrm{z}}(\nu,\tau) &= - M_{\mathrm{s}} B_0 \int_V m_z \, d^3 r =- \frac{4\pi R^3 M_{\mathrm{s}} B_0}{3} \, g_{\mathrm{z}}^{z}(\nu)\cos\tau , \\
\mathcal{H}_{\mathrm{d}}(\nu) &=- \frac{1}{2} M_{\mathrm{s}} \int_V \mathbf{m}\cdot\mathbf{B}_{\mathrm{d}} \, d^3 r =- \frac{4\pi R^3 \mu_0 M_{\mathrm{s}}^2}{3} \, g_{\mathrm{d}}(\nu) ,
\end{align}
where the energy profile integrals are defined as follows:
\begin{align}
g_{\mathrm{ex}}(\nu) &= \frac{3}{2}\int_{0}^{\pi}\int_{0}^{1} \frac{\nu^2 \xi^2 \sin^2\theta + \sinh^2(\nu \xi \sin\theta)}{\sin\theta \cosh^2(\nu \xi \sin\theta)} \, d\xi \, d\theta, \label{eq:VortexModelIntegralProfile1}\\
g_{\mathrm{u}}^{x}(\nu) &= \frac{3}{4} \int_{0}^{\pi} \int_{0}^{1} \tanh^2(\nu \xi \sin\theta) \, \xi^2 \sin\theta \, d\xi \, d\theta, \label{eq:VortexModelIntegralProfile2}\\
g_{\mathrm{u}}^{z}(\nu) &= \frac{3}{2} \int_{0}^{\pi} \int_{0}^{1} \operatorname{sech}^2(\nu \xi \sin\theta) \, \xi^2 \sin\theta \, d\xi \, d\theta, \label{eq:VortexModelIntegralProfile3}\\
g_{\mathrm{z}}^{z}(\nu) &= \frac{3}{2} \int_{0}^{\pi} \int_{0}^{1} \operatorname{sech}(\nu \xi \sin\theta) \, \xi^2 \sin\theta \, d\xi \, d\theta. \label{eq:VortexModelIntegralProfile4}
\end{align}
A normalized spherical coordinate system $(\xi, \theta, \phi)$ is employed, with $0 \le \xi \le 1$, where the relations to the previously used cylindrical coordinates $(\rho, \phi, z)$ are given by $\rho/R = \xi \sin\theta$ and $\tan\theta = \rho/z$. Due to the rotational symmetry of both the exchange and magnetodipolar interactions, and the isotropy of the spherical geometry, these energy terms remain invariant under the transformation defined by the rotation matrix~\eqref{eq:VortexParticlesRotationMatrixapp}. The integral expression for the exchange profile function $g_{\mathrm{ex}}$ was obtained by explicitly expressing the Cartesian gradients $\nabla m_i$ in cylindrical coordinates using the chain rule and subsequently expressing the volume integral in normalized spherical coordinates. The integration over the $\phi$~coordinate is trivial due to rotational symmetry.

The dipolar energy contribution corresponds to a six-dimensional integral over the magnetization vector field, reflecting the long-range and nonlocal character of the dipole-dipole interaction. This integral is evaluated numerically by discretizing the unit sphere using a regular Cartesian grid with spacing $\delta$ and volume element $\Delta V$ (compare to Refs.~\cite{evans2014vampire,johnston2016magnetic,muller2019spirit}):
\begin{align}
&g_{\mathrm{d}}(\nu) = \frac{3}{4\pi}\frac{\Delta V}{2\mu_0} \sum_{i=1}^{n} \mathbf{m}'(\nu \xi_i, \theta_i ,\phi_i) \cdot \mathbf{b}_{\mathrm{d}}(\xi_i, \theta_i, \phi_i),\label{eq:VortexModelIntegralProfile5}
\end{align}
where the reduced discrete demagnetizing field is computed as
\begin{align}
&\mathbf{b}_{\mathrm{d}}(\xi_i, \theta_i, \phi_i) = \frac{2\mu_0}{3} \mathbf{m}'(\nu \xi_i, \theta_i, \phi_i) \\
&+ \frac{\mu_0 \Delta V}{4\pi} \sum_{\substack{j=1\\ j \neq i}}^{n} \left[
3\frac{(\boldsymbol{\xi}_{i} - \boldsymbol{\xi}_j) \otimes (\boldsymbol{\xi}_{i} - \boldsymbol{\xi}_j)}{\|\boldsymbol{\xi}_{i} - \boldsymbol{\xi}_j\|^5}
- \frac{\mathbf{I}}{\|\boldsymbol{\xi}_{i} - \boldsymbol{\xi}_j\|^3}
\right] \cdot \mathbf{m}'(\nu \xi_j, \theta_j, \phi_j). \nonumber
\end{align}
More specifically, the discrete positions $\boldsymbol{\xi}_i$ used in the evaluation are defined on a uniform cubic grid as
\begin{align}
    \boldsymbol{\xi}_i = (k_i^{x}\delta -1)\mathbf{e}_x + (k_i^{y}\delta -1)\mathbf{e}_y + (k_i^{z}\delta -1)\mathbf{e}_z, \quad \|\boldsymbol{\xi}_i\| \le 1, 
\end{align}
where $k_i^{x}, k_i^{y}, k_i^{z} \in \{0,\ldots,K-1\}$ are integer indices along the Cartesian axes, $\delta = 2/(K-1)$ is the step size, and $K$ denotes the number of grid points per spatial direction. The volume element of the regular Cartesian grid is given by $\Delta V = \delta^3$. The results presented in the following were obtained using $K = 21$ grid points per spatial direction, corresponding to a total of $\mathcal{N} = 4169$ grid points within the unit sphere.

\begin{figure}[tb!]
\centering
\resizebox{0.520\columnwidth}{!}{\includegraphics{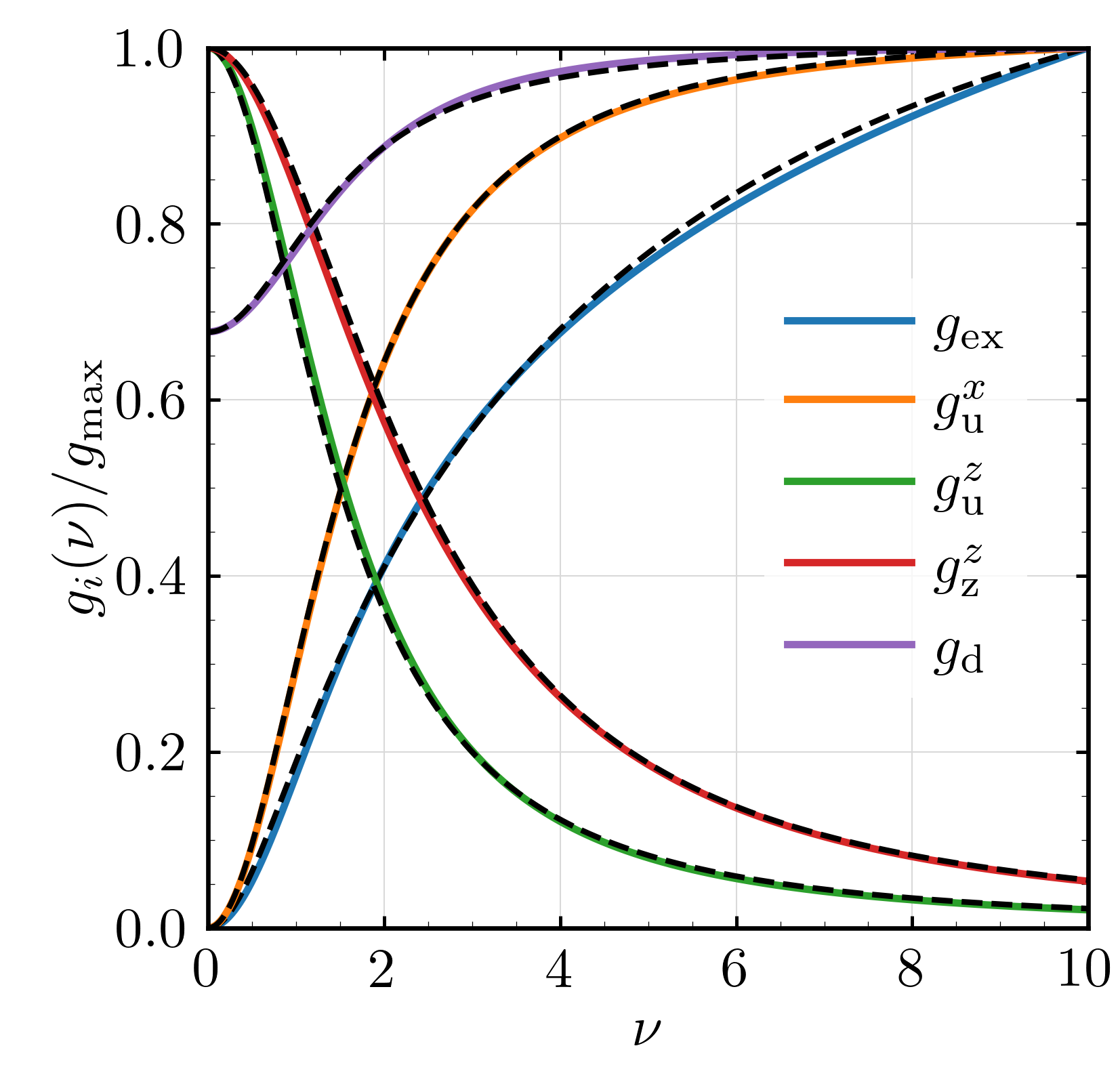}}
\caption{Comparison between the numerically evaluated dimensionless profile functions \( g_i(\nu)>0 \) defined by the integral expressions [Eqs.~\eqref{eq:VortexModelIntegralProfile1}$-$\eqref{eq:VortexModelIntegralProfile5}] (solid colored lines) and their respective closed-form heuristic approximations [Eqs.~\eqref{eq:HeuristicProfile1}$-$\eqref{eq:HeuristicProfile5}] (dashed black lines), normalized by their individual maxima \( g_i^\mathrm{max} \). The approximations reproduce the behavior of the original integrals with high fidelity in the relevant domain \( \nu \in [0, 10] \), while maintaining minimal analytical complexity and smooth asymptotic behavior. Except for the exchange profile \( g_{\mathrm{ex}} \), all \( g_i(\nu) \) appear with a negative sign in the Hamiltonian \( \mathcal{H}' \) [Eq.~\eqref{eq:VortexReducedHamiltonianAppx}], so that increasing \( g_i(\nu) \) does not necessarily imply an increase in energy.}
\label{fig1sm}
\end{figure}

Finally, the reduced total energy is expressed in terms of the profile functions $g_{i}(\nu)$ as:
\begin{align}
    \mathcal{H}'(\nu, \tau) &= g_{\mathrm{ex}}(\nu) - \frac{K_{\mathrm{u}} R^2}{A}  \left(g_{\mathrm{u}}^{x}(\nu) \sin^2\tau + g_{\mathrm{u}}^{z}(\nu) \cos^2\tau\right) 
    \label{eq:VortexReducedHamiltonianAppx}
    \\
    &- \frac{M_{\mathrm{s}} R^2 B_0}{A} g_{\mathrm{z}}^{z}(\nu) \cos\tau -\frac{\mu_0 M_{\mathrm{s}}^2 R^2}{A} g_{\mathrm{d}}(\nu) \nonumber .
\end{align}
To enable a more efficient analytical treatment, we propose the following set of closed-form heuristic approximations for the profile functions $g_i(\nu)$ defined by the integral expressions Eqs.~\eqref{eq:VortexModelIntegralProfile1}$-$\eqref{eq:VortexModelIntegralProfile5}. These are valid for the range $0 \le \nu \le 10$.
\begin{align}
g_{\mathrm{ex}}(\nu) &\cong 1.95 \; \log(1 + 159 \; \tanh^2(\nu/11)), \label{eq:HeuristicProfile1}
\\
g_{\mathrm{u}}^{x}(\nu) &\cong 1.32 \frac{(\nu/1.551)^2 + (\nu/3.642)^4}{1 + (\nu/1.551)^2 + (\nu/3.642)^4}, \label{eq:HeuristicProfile2}
\\
g_{\mathrm{u}}^{z}(\nu) &\cong \frac{1}{1 + (\nu/1.5)^2}, \label{eq:HeuristicProfile3}
\\
g_{\mathrm{z}}^{z}(\nu) &\cong \frac{1}{1 + (\nu/2.4)^2}, \label{eq:HeuristicProfile4}
\\
g_{\mathrm{d}}(\nu) &\cong 0.332 + 0.162\frac{(\nu/1.512)^2}{1 + (\nu/1.512)^2}.\label{eq:HeuristicProfile5}
\end{align}
These approximations are not derived from an analytical evaluation of the integrals, but are constructed heuristically to minimize the number of fit parameters while ensuring a high quantitative agreement in the physically relevant range. A comparison between the original integrals and the approximations is shown in Fig.~\ref{fig1sm}.

The physical interpretation of the profile functions is as follows: All profile functions shown in Fig.~\ref{fig1sm} are positive-valued (\(g_i(\nu) > 0\)), but, except for the exchange contribution, they enter the Hamiltonian \(\mathcal{H}'\) with a negative sign [compare Eq.~\eqref{eq:VortexReducedHamiltonianAppx}]. Consequently, an increasing \(g_i(\nu)\) does not necessarily correspond to a larger energy contribution. This must be kept in mind when interpreting the plots: for most terms, larger values of \(g_i(\nu)\) actually imply a stronger reduction of the corresponding energy during the transition from saturation to remanence.

As $\nu$ increases, the system transitions from a field-polarized state (uniform) to a vortex-like configuration comparable to the one shown in Fig.~1(a) in the main text. During this process, the exchange energy increases due to decreasing spin alignment, while the demagnetization energy decreases as a result of improved flux closure. The Zeeman energy increases with stronger deviation from the external field direction. The out-of-plane anisotropy contribution $g_{\mathrm{u}}^z$ increases the anisotropy energy, whereas the in-plane component $g_{\mathrm{u}}^x$ leads to a reduction. The sign of each contribution is consistent with the structure of the Hamiltonian.

It should be noted that the functional form chosen for the exchange profile $g_{\mathrm{ex}}(\nu)$ differs qualitatively from the other approximations, as it does not exhibit saturation within the range considered. This reflects the physical nature of the exchange interaction, which penalizes spatial gradients in the magnetization. As $\nu$ increases, the vortex becomes more localized and the spin alignment degrades, especially near the vortex core. This leads to a continuous growth of the exchange energy, in contrast to the saturating behavior of the other contributions. The logarithmic form used in Eq.~\eqref{eq:HeuristicProfile1} captures this gradual but unbounded increase, while preserving analytic tractability.

\section{Derivation of the critical vortex-nucleation size}
\label{appb}

An instructive benchmark for the present modeling approach is the prediction of the critical vortex-nucleation size $R_{\mathrm{nuc}}$, which characterizes the competition between exchange energy, anisotropy, and dipole-dipole interactions. Since the vortex-core parameter $\nu = 0$ corresponds to the transition point between a uniform magnetization state and a vortex-like state, the stability of the homogeneous configuration is analyzed by evaluating the curvature of the following reduced (minimal) Hamiltonian with respect to $\nu$.
\begin{align}
\label{eq:VortexReducedSimplifiedHamiltonian}
    \mathcal{H}''(\nu, \tau) &= g_{\mathrm{ex}}(\nu) - \frac{K_{\mathrm{u}} R^2}{A}  g_{\mathrm{u}}^{z}(\nu) \cos^2\tau \\
    &- \frac{M_{\mathrm{s}} R^2 B_0}{A} g_{\mathrm{z}}^{z}(\nu) \cos\tau -\frac{\mu_0 M_{\mathrm{s}}^2 R^2}{A} g_{\mathrm{d}}(\nu) \nonumber .
\end{align}
Further, we select the inclination angle $\tau=0$, which corresponds to the saturated
regime for a positive applied magnetic field $B_0$ (see Fig.~3 in the main paper). The first derivative vanishes:
\begin{align}
    \left.\frac{\partial \mathcal{H}''}{\partial \nu } \right|_{\nu = 0, \; \tau = 0} &= 0,
\end{align}
and the second derivative determines the local stability of the single-domain state:
\begin{align}
    \left.\frac{\partial^2 \mathcal{H}''}{\partial \nu^2 } \right|_{\nu = 0,\; \tau = 0} 
    =   4 
    + \frac{4}{5}\frac{K_{\mathrm{u}} R^2}{A} 
    + \frac{2}{5}\frac{B_0 M_{\mathrm{s}} R^2}{A} 
    - \frac{125}{882} \frac{\mu_0 M_{\mathrm{s}}^2 R^2}{A}  .
    \label{eq:StabilityConditionAppx}
\end{align}
We note that the prefactors of the exchange, anisotropy, and Zeeman contributions are exact results obtained from the integrals Eqs.~\eqref{eq:VortexModelIntegralProfile1}$-$\eqref{eq:VortexModelIntegralProfile4}, whereas the prefactor of the dipole-dipole contribution is taken from the effective quadratic coefficient of the corresponding profile function [Eq.~\eqref{eq:HeuristicProfile5}], since the full magnetostatic energy integral cannot be evaluated analytically.

The critical vortex-nucleation radius \( R_{\mathrm{nuc}} \) follows from the condition that this expression vanishes. In the special case \( B_0 = 0 \), the zero-field instability threshold takes on the form (for $K_{\mathrm{u}} \lesssim 0.177 \mu_0 M_{\mathrm{s}}^2$):
\begin{align}
\label{eq:CriticalVortexRadiusAdams}
    R_{\mathrm{nuc}} 
    &=\frac{42\sqrt{5}}{25} \ell_{\mathrm{ex}}\left[1 - \frac{3528}{625}\frac{K_{\mathrm{u}}}{\mu_0 M_{\mathrm{s}}^2}\right]^{-1/2},
    \\
    &\cong 3.76 \, \ell_{\text{ex}} \left[1 - 5.64   \frac{K_{\mathrm{u}}}{\mu_0 M_{\mathrm{s}}^2}\right]^{-1/2} \nonumber ,
\end{align}
where the exchange length $\ell_{\mathrm{ex}}$ is defined as 
\begin{align}
    \ell_{\mathrm{ex}} = \sqrt{\frac{2A}{\mu_0 M_{\mathrm{s}}^2}} .
\end{align}
Note that $\ell_{\mathrm{ex}} \cong 3$$-$$10 \, \mathrm{nm}$ for many magnetic materials (see Table~3.1 in \cite{kronmuller2003micromagnetism}). The analytically derived expression~\eqref{eq:CriticalVortexRadiusAdams} for the critical particle radius \( R_{\mathrm{nuc}} \) at which the homogeneous magnetization becomes unstable toward a vortex-like distortion is formally equivalent in structure to the classic result of Brown~\cite{brown1968fundamental,aharoni2000introduction}:
\begin{align}
    R_{\mathrm{nuc}}^{\mathrm{Brown}} \cong 4.53 \, \ell_{\mathrm{ex}} \left[1 - 5.62   \frac{K_{\mathrm{u}}}{\mu_0 M_{\mathrm{s}}^2}\right]^{-1/2}.\label{eq:CriticalVortexRadiusBrown}
\end{align}
Note that Brown's original notation uses the constants $C = 2A$, $\gamma = \mu_0$, and $K_1 = K_{\mathrm{u}}$. Although the numerical prefactors in Eqs.~\eqref{eq:CriticalVortexRadiusAdams} and~\eqref{eq:CriticalVortexRadiusBrown} differ slightly, their functional forms are identical. The remaining numerical discrepancy is within an acceptable range and can be attributed to the differing assumptions underlying the respective models and Ansatz functions.

Brown derived his approximation based on a static curling-type magnetization field (compare to Eq.~\eqref{ansatz1Appx} and Refs.~\cite{brown1968fundamental,aharoni2000introduction}),
\begin{align}
    \mathbf{m}_{\mathrm{Brown}}'(\rho, \phi, z)
    =
    \left\{
    -\sqrt{1 - m_z^2}\sin\phi,\;
    +\sqrt{1 - m_z^2}\cos\phi,\;
    1 - \frac{\rho^2}{R^2}
    \right\}.
\end{align}
While Brown's calculation is based on a fixed curling-type magnetization field, the present approach generalizes this idea by introducing a variational degree of freedom $\nu$, which continuously interpolates between the uniform state ($\nu = 0$) and a vortex-like configuration ($\nu \gg 1$). For a direct comparison with Brown's classic result, the expression in Eq.~\eqref{eq:CriticalVortexRadiusAdams} is evaluated under the zero-field condition $B_0 = 0$. At the same time, the generalized framework developed here naturally accommodates arbitrary applied fields. As a consequence, the stability condition~\eqref{eq:StabilityConditionAppx} can be solved explicitly for the applied field, yielding a corresponding vortex-nucleation field
\begin{align}
    B_{\mathrm{nuc}} =
    \frac{625}{1764} \mu_0 M_{\mathrm{s}}
    - \frac{2 K_{\mathrm{u}}}{M_{\mathrm{s}}}
    - \frac{10A}{M_{\mathrm{s}} R^2}.
\end{align}
Notably, the second term, $2K_{\mathrm{u}}/M_{\mathrm{s}}$, coincides with the Stoner-Wohlfarth switching field, highlighting that vortex nucleation occurs relative to the loss of stability of the uniform macrospin state (see below).

\section{Stoner-Wohlfarth limit}
\label{appc}

\begin{figure}[tb!]
\centering
\resizebox{0.55\columnwidth}{!}{\includegraphics{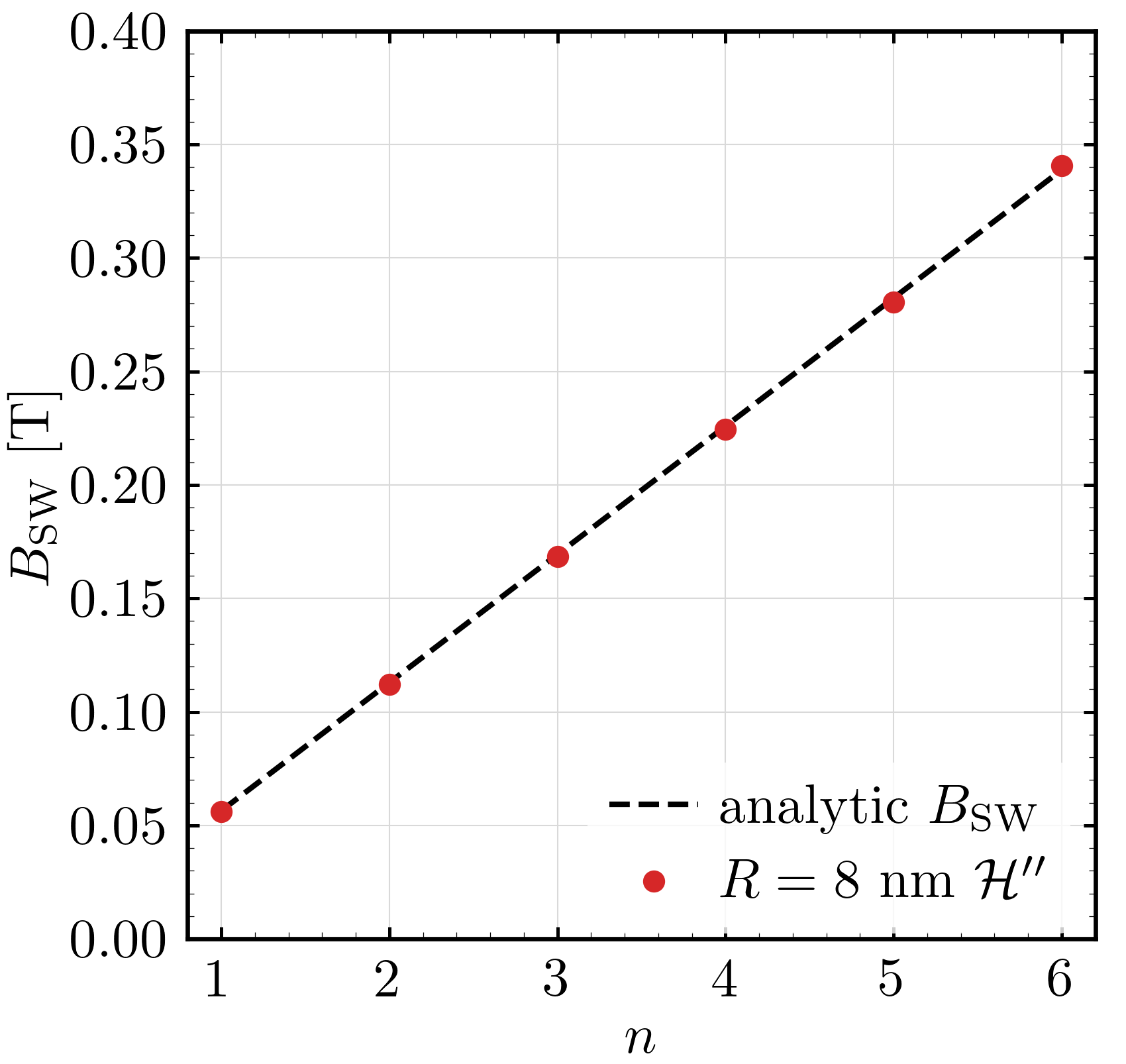}}
\caption{Stoner-Wohlfarth switching field $B_{\mathrm{SW}}$ as a function of the anisotropy scaling factor $n$ in $K_{\mathrm{u}} \rightarrow n K_{\mathrm{u}}$. ($\color{red}\bullet$)~$B_{\mathrm{SW}}$ obtained from the minimal vortex Hamiltonian $\mathcal{H}''$ [Eq.~\eqref{eq:VortexReducedSimplifiedHamiltonian}]. (Dashed line) $B_{\mathrm{SW}} = 2 n K_{\mathrm{u}}/M_{\mathrm{s}}$. For the chosen materials parameters of iron, $R_{\mathrm{nuc}} \cong 9.2 \, \mathrm{nm}$ and $B^{n=1}_{\mathrm{SW}} \cong 0.056 \, \mathrm{T}$.}
\label{fig2sm}
\end{figure}

The Stoner-Wohlfarth limit corresponds to the uniform magnetization configuration obtained for $\nu = 0$, for which the hyperbolic vortex ansatz [Eq.~\eqref{ansatz1Appx}] reduces to $\mathbf{m}' = \mathbf{e}_z$. In this limit, the vortex degree of freedom is suppressed and $\mathcal{H}''$ simplifies to
\begin{align}
    \mathcal{H}''(\nu=0, \tau, B_0) &=
    - \frac{K_{\mathrm{u}} R^2}{A} \cos^2\tau
    - \frac{M_{\mathrm{s}} R^2 B_0}{A}\cos\tau
    - 0.332\,\frac{\mu_0 M_{\mathrm{s}}^2 R^2}{A} .
\end{align}
Here, the exchange contribution vanishes identically, while the demagnetization energy reduces to a constant offset. The remaining energy landscape is therefore governed by the uniaxial anisotropy and the Zeeman interaction, both depending on the magnetization angle~$\tau$. Up to an overall prefactor, this expression is identical to the well-known uniaxial macrospin energy with the easy axis aligned along the external field. The switching field follows from the loss of local stability of the metastable states along the hysteresis loop. Since the energy landscape is symmetric with respect to $\tau = 0$ and $\tau = \pi$, the loss of local stability occurs symmetrically at both orientations, $\tau \in \{0,\pi\}$. This instability is determined by the condition
\begin{equation}
\left.\frac{\partial^2 \mathcal{H}''}{\partial \tau^2}\right|_{\tau\in \{0,\pi\}} = 0,
\end{equation}
which yields the Stoner-Wohlfarth switching field
\begin{equation}
\label{hswtheo}
B_{\mathrm{SW}} = \mu_0 H_{\mathrm{SW}} =  \pm \frac{2K_{\mathrm{u}}}{M_{\mathrm{s}}}.
\end{equation}
Hence, in the single-domain limit, the present effective Hamiltonian reproduces the exact Stoner-Wohlfarth result analytically (as it should be). Deviations from coherent rotation for larger particle sizes arise solely from the activation of the vortex degree of freedom and do not require any modification of the underlying theoretical framework. Figure~\ref{fig2sm} shows that the Stoner-Wohlfarth switching field obtained from micromagnetic simulations using $\mathcal{H}''$ perfectly agrees with the theoretical $B_{\mathrm{SW}} = 2 n K_{\mathrm{u}}/M_{\mathrm{s}}$ (dashed line), where $n$ denotes the anisotropy scaling factor ($K_{\mathrm{u}} \rightarrow n K_{\mathrm{u}}$).

\section{Effect of mesh size in \textsc{MuMax3} simulations}
\label{appd}

\begin{figure}[tb!]
\centering
\resizebox{0.55\columnwidth}{!}{\includegraphics{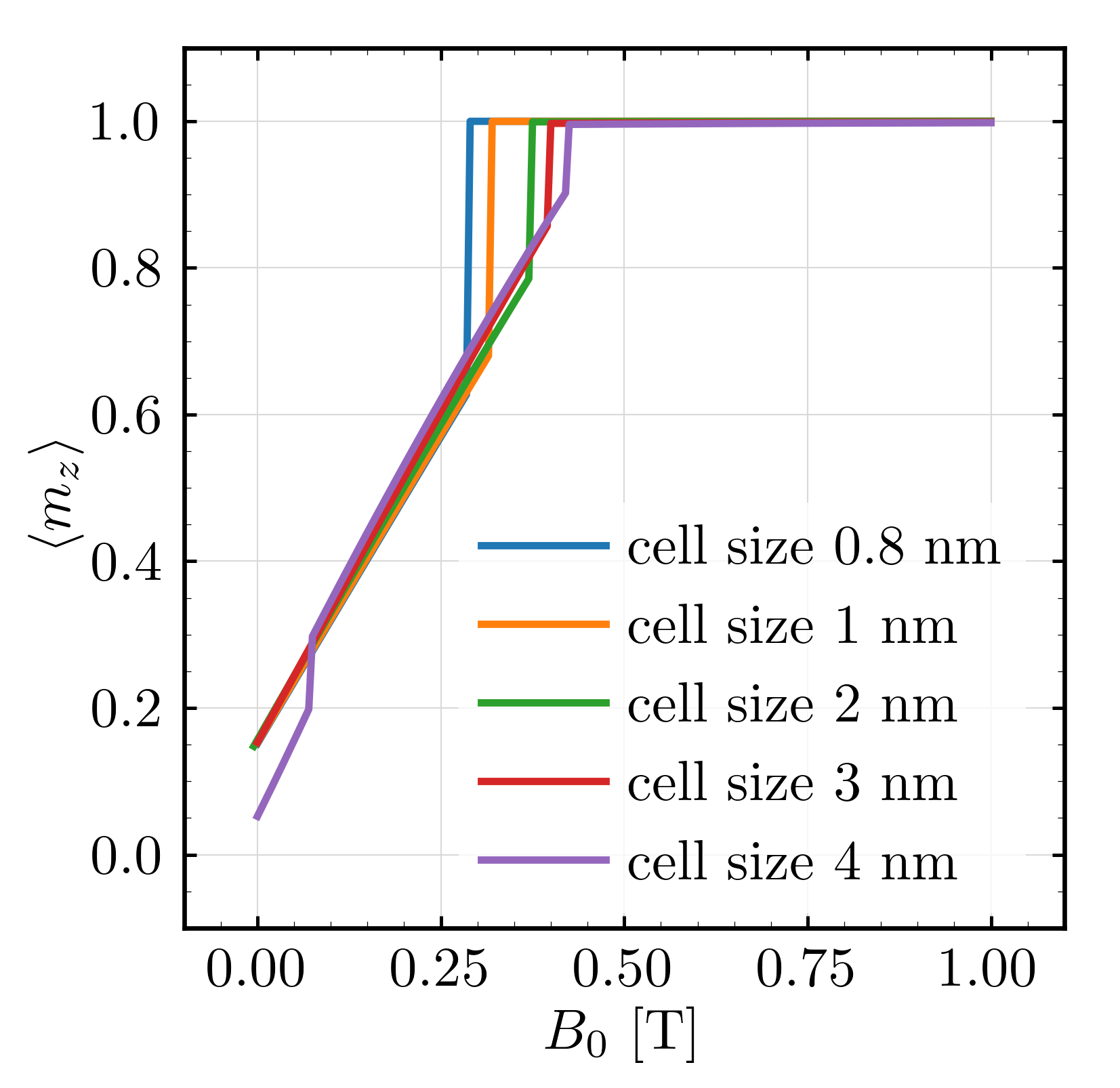}}
\caption{Field-dependent magnetization curves obtained from \textsc{MuMax3} simulations for varying size of the discretization cell (see inset).}
\label{fig3sm}
\end{figure}

It should be emphasized that micromagnetic simulations are themselves modeling approaches that should be interpreted with care. As shown in Fig.~\ref{fig3sm}, both the remanence and the upper-loop switching characteristics of the magnetization exhibit a clear dependence on the size of the discretization mesh used. The observed discontinuities in the hysteresis curve---particularly during field reduction from saturation---may originate from mesh-induced symmetry breaking. In a true continuum description, such abrupt switching events might instead appear as smoother transitions. Clarifying the origin and nature of this behavior would require further investigation and would constitute a nontrivial problem in its own right.

{\it Data availability}---Numerical data and codes (MuMax3 and python scripts) corresponding to this study are available on Zenodo~\cite{Adams2026_Vortex}.

\bibliography{references}